\documentclass[superscriptaddress,preprint]{revtex4}
\usepackage{mathrsfs}
\usepackage{mathrsfs}
\usepackage{mathrsfs}
\usepackage{mathrsfs}
\usepackage{amssymb}
\usepackage[tbtags]{amsmath}
\usepackage{graphicx}
\usepackage{epsfig,graphicx,times}
\usepackage{color}
\usepackage{subfigure}

\begin{document}
\title{Spin-orbit couplings between distant electrons
trapped individually on liquid Helium}
\author{M. Zhang}
\affiliation{Quantum Optoelectronics Laboratory, School of Physics,
Southwest Jiaotong University, Chengdu 610031, China}
\author{L. F. Wei\footnote{weilianfu@gmail.com}}
\affiliation{Quantum Optoelectronics Laboratory, School of Physics,
Southwest Jiaotong University, Chengdu 610031, China}
\affiliation{State Key Laboratory of Optoelectronic Materials and
Technologies, School of Physics and Engineering, Sun Yat-sen
University, Guangzhou 510275, China}
\date{\today}

\begin{abstract}
We propose an approach to entangle spins of electrons floating on
the liquid Helium by coherently manipulating their spin-orbit
interactions. The configuration consists of single electrons,
confined individually on liquid Helium by the micro-electrodes,
moving along the surface as the harmonic oscillators. It has been
known that the spin of an electron could be coupled to its orbit
(i.e., the vibrational motion) by properly applying a magnetic
field. Based on this single electron spin-orbit coupling, here we
show that a Jaynes-Cummings (JC) type interaction between the spin
of an electron and the orbit of another electron at a distance could
be realized via the strong Coulomb interaction between the
electrons. Consequently, the proposed JC interaction could be
utilized to realize a strong orbit-mediated spin-spin coupling and
implement the desirable quantum information processing between the
distant electrons trapped individually on liquid Helium.

\end{abstract}

\maketitle

\section{Introduction}
The interactions between the microscopic particles, e.g., the ions
in Paul trap~\cite{Cirac}, the neutral atoms confined in optical
lattice~\cite{Optical lattice}, and the electrons in Penning
trap~\cite{Penning trap}, etc., relate usually to their masses and
the inter-particle forces. Due to the small mass and the strong
Coulomb interaction, the interacting electrons could be used to
implement quantum information processing (QIP). The idea of quantum
computing with strongly-interacting electrons on liquid Helium was
first proposed by Platzman and Dykman in 1999~\cite{Science}. In
their proposal, the two lower hydrogen-like levels of the
surface-state electron are encoded as a qubit, and the effectively
interbit couplings can be realized by the electric dipole-dipole
interaction. When the liquid helium is cooled on the order of mK
temperature the qubit possesses long coherent time (e.g., up to the
order of ms)~\cite{PRB,Microwave Saturation}. Interestingly, Lyon
suggested~\cite{Spin} that the qubits could also be encoded by the
spins of the electrons on liquid Helium, and estimated that the
qubit coherent time could reach $100$~s~\cite{Spin}. He showed
further that the magnetic dipole-dipole interactions between the
spins could be used to couple the qubits, if the electrons are
confined closed enough. For example, the coupling strength can reach
to the order of kHz for the distance $d=0.1\,\mu$m between the
electrons~\cite{Spin}. Remarkably, recent
experiments~\cite{experiment-1,experiment-2,experiment-3}
demonstrated the manipulations of electrons (confining,
transporting, and detecting) on liquid Helium in the single-electron
regime. This provides really the experimental platforms to realize
the relevant QIP with electrons on liquid
Helium~\cite{Miao-e,Miao-OL,electronQED,Simulator}.

Here, we propose an alternative approach to implement QIP with
electronic spins on liquid Helium by coherently manipulating the
spin-orbit interactions of the electrons. In our proposal, the
virtues of long-lived spin states (to encode the qubit) and strong
Coulomb interaction (for realizing the expectably-fast interbit
operations) are both utilized. The electrons are trapped
individually on the surface of liquid Helium by the
micro-electrodes. In the plane of liquid Helium surface each
electron moves as a harmonic oscillator. It has been showed that
such an external orbit-vibration could be effectively coupled to the
internal spin of a single electron by applying a magnetic field with
a gradient along the vibrational axis~\cite{electronQED}.
Interestingly, we show that the spin of an electron could be coupled
to the vibrational motion of another distant electron [as a
Jaynes-Cummings (JC) type interaction], by designing a proper
virtual excitation of the electronic vibration. The present JC
interaction could be utilized to significantly enhance the spin-spin
coupling between the distant electrons, and implement the desirable
quantum computation with the spin qubits on liquid Helium.

The paper is organized as follows: In Sec.~II we discuss the
mechanism for spin-orbit coupling with a single electron trapped on
liquid Helium~\cite{electronQED}, and then show how to utilize such
a coupling to realize the desirable quantum gate with the single
electron. By using the electron-electron Coulomb interaction, in
Sec.~III, we propose an approach to implement the JC coupling
between the spin of an electron and the orbital motion of another
electron. Based on such a distant spin-orbit interaction, we show
that a two-qubit controlled-NOT (CNOT) gate and an orbit-enhanced
coupling between the distant spins could be implemented. Finally, we
give a conclusion in Sec. IV.

\begin{figure}[tbp]
\includegraphics[width=7.5cm]{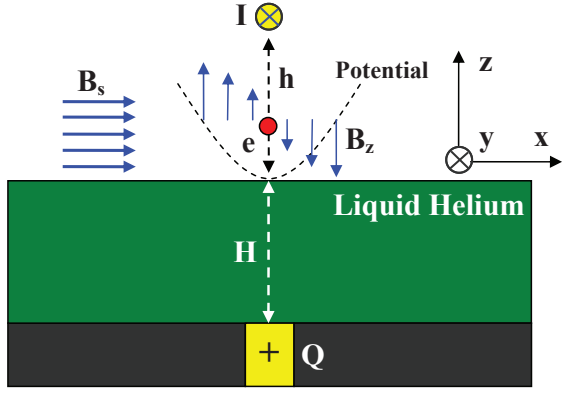}
\caption{(Color online) Sketch for a single electron trapped on the
surface of liquid Helium. The liquid Helium provides $z$-directional
confinement, and the micro-electrode Q (below the Helium surface at
depth $H$) traps the electron in $x$-$y$ plane. The desirable spin
qubit is generated by an applied uniform magnetic field $B_s$, and
the spin-orbit coupling of the trapped electron is obtained by
applying a current to another micro-electrode $\text{I}$ (upon the
liquid Helium surface at the height $h$).}
\end{figure}

\section{Spin-orbit coupling with a single trapped electron}
We consider first a single electron trap shown in
Fig.~1~\cite{electronQED}, wherein an electron (with mass $m_e$ and
charge $e$) on liquid Helium is weakly attracted by its dielectric
image potential $V(z)=-\Lambda e^2/z$ (with
$\Lambda=(\varepsilon-1)/4(\varepsilon+1)$ and $\varepsilon$ being
the dielectric constant of liquid Helium). Due to the Pauli
exclusion principle, there is an barrier (about $1$~eV) to prevent
the electron penetrating into the liquid Helium.
As a consequence, $z$-directional confinement of the electron is
realized, yielding an one-dimensional (1D) hydrogenlike atom with
the spectrum $E_n=-\hbar R/n^2$~\cite{Girmes}. Here,
$R=\Lambda^2e^4m_e/(2\hbar^2)\approx170$~GHz and
$r_{\rm{b}}=\hbar^2/(m_ee^2\Lambda)\approx7.6$~nm are the effective
Rydberg energy and Bohr radius, respectively.
In $x$-$y$ plane, the electron can be confined by the
micro-electrode Q located at $H$ beneath the liquid Helium surface.
Typically, $x,y,z\ll H$, and thus the potential of the electron can
be described by~\cite{PRB}
\begin{equation}
U(x,y,z)\approx-\frac{\Lambda e^2}{z}+E_\perp z
+\frac{m_e}{2}(\nu_x^2x^2+\nu_y^2y^2)
\end{equation}
with $E_\perp=eQ/H^2$, $\nu_x=\nu_y=\sqrt{eQ/(m_eH^3)}$, and $Q$
being the effective charge of the micro-electrode. This potential
indicates that the motions of the trapped electron are a 1D
Stark-shifted hydrogen along the $z$-direction, and a 2D harmonic
oscillator in the plane parallel to the liquid Helium surface.
The Hamiltonian for the orbital motions of the trapped electron can
be written as
\begin{equation}
\hat{H}_{o}=\sum_nE_n|n_a\rangle\langle
n_a|+\sum_{k=x,y}\hbar\nu_k(\hat{a}_k^\dagger\hat{a}_k+\frac{1}{2}).
\end{equation}
Here, $|n_a\rangle$ is the $n$th bound state of the hydrogenlike
atom, $\hat{a}_k^\dagger$ and $\hat{a}_k$ are the bosonic operators
of the vibrational quanta of the electron along the $k$-direction.

A spin qubit is generated by applying an uniform magnetic field
$B_s$ along $x$ direction, and its Hamiltonian reads
$\hat{H}_{q}=(g\mu_BB_s)\hat{\sigma}_x/2 $. Here, the Pauli operator
is defined as $\hat{\sigma}_x=|\uparrow\rangle\langle\uparrow|
-|\downarrow\rangle\langle\downarrow|$ with $|\downarrow\rangle$ and
$|\uparrow\rangle$ being the two spin states. $g=2$ is the
electronic $g$-factor, and $\mu_B = 9.3\times10^{-24}$~J/T is the
Bohr magneton. The spin-orbit coupling of the trapped electron can
be realized by applying a dc current $I$ to the electrode $\text{I}$
(located upon the liquid Helium surface with a height
$h$)~\cite{electronQED}. Typically, $x,z\ll h$ and the magnetic
field generated by the current $I$ reads $\vec{B}=(B_x,0,B_z)$ with
$B_x\approx\mu_0I(1-z/h)/(2\pi h)$ and $B_z\approx\mu_0Ix/(2\pi
h^2)$. Here, $\mu_0$ is the permeability of free space. Therefore,
the Hamiltonian describing the interaction between the magnetic
field and spin can be expressed as:
$\hat{H}_{\rm{sb}}=g\mu_B(B_z\hat{\sigma}_z+ B'_x\hat{\sigma}_x)/2$
with $B'_x=B_s+B_x$, $\hat{\sigma}_z=\hat{\sigma}_-
+\hat{\sigma}_+$,
$\hat{\sigma}_-=|\downarrow\rangle\langle\uparrow|$ and
$\hat{\sigma}_+=|\uparrow\rangle\langle\downarrow|$.
Consequently, the total Hamiltonian of the trapped electron in the
applied magnetic fields reads
\begin{equation}
\hat{H}=\frac{\hbar\nu_s}{2}\hat{\sigma}_x+\hat{H}_o+\hat{H}_{sx},
\end{equation}
with
\begin{equation}
\hat{H}_{sx}=
\frac{g\mu_B\mu_0I}{4\pi
h^2}\sqrt{\frac{\hbar}{2m_e\nu_x}}\,(\hat{a}_x+\hat{a}_x^\dagger)
\hat{\sigma}_z.
\end{equation}
The first and second terms in the right hand of Eq.~(3) describe the
free Hamiltonian of the trapped electron, with
$\nu_s=(g\mu_B/\hbar)[B_s+(\mu_0I/2\pi h)]$ being the transition
frequency between its two spin states, and $\hat{H}_{sx}$ describes
the coupling between the spin and the orbital motion along
$x$-direction. Note that the coupling between the spin and
$z$-directional orbital motion is neglected, due to the
large-detuning. Also, the applied strong field $B_s$ (e.g.,
$0.06$~T) does not affect the interaction $\hat{H}_{sx}$, although
it will change slightly the electron's motions in the $y$-$z$
plane~\cite{EIB}.

Obviously, the Hamiltonian in Eq.~(3) can be simplified as
\begin{equation}
\hat{H}_{e}=\hbar\Omega\left(e^{i\delta t}\hat{\sigma}_+\hat{a}
+e^{-i\delta t}\hat{\sigma}_-\hat{a}^\dagger\right)
\end{equation}
in the interaction picture. Here, $\delta=\nu_s-\nu_x$ is the
detuning,
\begin{equation}
\Omega=\frac{g\mu_B\mu_0I}{4\pi h^2\sqrt{2\hbar m_e\nu_x}}
\end{equation}
is the coupling strength, and
$\hat{a}=\hat{a}_x$,\,$\hat{a}^\dagger=\hat{a}_x^\dagger$. Note
that, the Hamiltonian in Eq.~(5) can also be obtained by applying an
ac current $I(t)=I\cos(\omega t)$ with frequency
$\omega=\nu_x-\nu_s+\delta$ to the electrode. Specially, when
$\delta=0$, this Hamiltonian describes a JC-type interaction between
the spin and orbit of the single electron. In fact,
Ref.~\cite{electronQED} has arranged this spin-orbit coupling of a
single electron to increase the interaction between the spin and a
quantized microwave field. Alternatively, we will utilize this
spin-orbit coupling (together with the electron-electron strong
Coulomb interaction) to realize a strong interaction between two
electronic spins and generate certain typical quantum gates.

For the typical parameters: $I=1$~mA, $h=0.5\,\mu$m, and
$\nu_x=10$~GHz~\cite{PRB,electronQED}, we have
$\Omega\approx5.2$~MHz. This is significantly larger than the
decoherence rate (which is typically on the order of
$10$~kHz~\cite{PRB,electronQED}) of the vibrational states of the
trapped electron. Thus, the above JC interaction provides a possible
approach to implement QIP between the spin and orbit states of a
single trapped electron. For the JC interaction, the
state-evolutions can be limited in the invariant-subspaces
$\{|\downarrow,0\rangle\}$ and
$\{|\downarrow,1\rangle,|\uparrow,0\rangle\}$, with $|0\rangle$ and
$|1\rangle$ being the ground and first excited states of the
harmonic oscillator. Thus, a phase gate
$\hat{P}=|0,\downarrow\rangle\langle0,\downarrow|
+|0,\uparrow\rangle\langle0,\uparrow|
+|1,\downarrow\rangle\langle1,\downarrow|-
|1,\uparrow\rangle\langle1,\uparrow|$ could be implemented by
applying a current pulse to the electrode  $\text{I}$. The relevant
duration $t$ is set to satisfy the conditions: $\sin(\Omega
t)\approx 0$ and $\cos(\sqrt{2}\Omega t)\approx-1$ (e.g., $\Omega
t\approx37.7$ numerically). Consequently, a CNOT gate with the
single electron could be realized as
$\hat{S}=\hat{R}(\pi/2,-\pi/2)\hat{P}\hat{R}(\pi/2,\pi/2)$, where
$\hat{R}(\alpha,\beta)=(|\uparrow\rangle\langle\uparrow|
+|\downarrow\rangle\langle\downarrow|)\cos(\alpha)
-i[\exp(i\beta)|\uparrow\rangle\langle\downarrow|
+\exp(-i\beta)|\downarrow\rangle\langle\uparrow|]\sin(\alpha)$ is an
arbitrary single-bit rotation~\cite{ESR}. This CNOT gate operation,
between the spin states and the two selected vibrational states of a
single electron~\cite{monroe}, is an intermediate step for the later
CNOT operation between two distant spin qubits.

\section{Spin-orbit JC coupling between the distant
electrons}Without loss of generality, we consider here two electrons
(denoted by $e_1$ and $e_2$) trapped individually in two potential
wells, see Fig.~2. Suppose that the distance $d$ between the
potential wells is sufficiently large (e.g., $d=10\,\mu$m), such
that the directly magnetic interaction between the two spins could
be neglected. Thus, the interaction between the two electrons leaves
only the Coulomb one. Specially, the Coulomb interaction along the
$x$-direction can be approximately written as
\begin{equation}
V(x)\approx\frac{e^2}{2\pi\epsilon_0d^3}x_1x_2
\end{equation}
with $x_j$ being the displacement of electron $e_j$ from its
potential minima. By controlling the voltages applied on the
electrodes $\rm{Q}_1$ and $\rm{Q}_2$, the vibrational frequencies of
the electrons are set as the large-detuning (and thus the electrons
are decoupled from each other).

\begin{figure}[tbp]
\includegraphics[width=7.5cm]{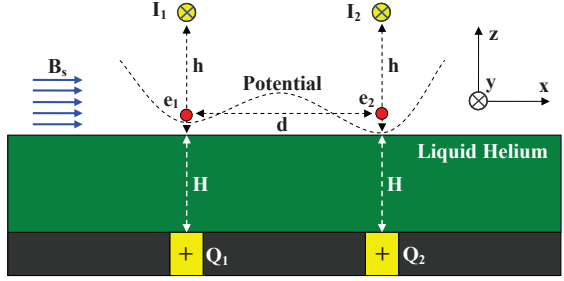}
\caption{(Color online) Two electrons (denoted by $e_1$ and $e_2$)
are confined individually in two potential wells with the distance
$d$, which is sufficiently large (e.g., $d=10\,\mu$m) such that the
magnetic dipole-dipole coupling between the electronic spins is
negligible. The orbital motions of the two electrons are also
decoupled from each other, since they are trapped in large-detuning
regime. By applying a current to the electrode $\text{I}_1$ the spin
of the electron $e_1$ could be coupled to the vibrational motions of
electron $e_2$, via a virtual excitation of the vibrational motion
of electron $e_1$.}
\end{figure}

To couple the initially-decoupled electrons, we apply a current $I$
to the electrode $\text{I}_1$. As discussed previously, such a
current induces a spin-orbit coupling [i.e., $\hat{H}_e$ in Eq.~(5)]
of the electron $e_1$. Therefore, the present two-electrons system
can be described by the following Hamiltonian
\begin{equation}
\hat{H}_{ee}=\hat{H}_{e}+\hbar \tilde{\Omega}\left(e^{i\Delta
t}\hat{a}\hat{b}^\dagger+e^{-i\Delta t}\hat{a}^\dagger\hat{b}\right)
\end{equation}
in the interaction picture. Where, $\hat{b}$ and $\hat{b}^\dagger$
are the bosonic operators of the vibrational motion of electron
$e_2$ along $x$-direction, $\Delta=\nu_{2x}-\nu_{1x}$ is the
detuning between the two electronic vibrations along $x$-direction,
and
\begin{equation}
\tilde{\Omega}=\frac{e^2}{4\pi\epsilon_0m_ed^3\sqrt{\nu_{1x}\nu_{2x}}}\,,
\end{equation}
the coupling strength.
Numerically, for $d=10\,\mu$m and $\nu_{jx}=10$~GHz we have
$\tilde{\Omega}\approx25$~MHz. Above, the spin of electron $e_2$ was
dropped, as the driving (induced by electrode $\rm{I}_1$) on this
spin is negligible (due to $d\gg h$).

The dynamical evolution ruled by the Hamiltonian in Eq.~(8) is given
by the following time-evolution operator
\begin{equation}
\begin{array}{l}
\hat{U}(t)=
1+\left(\frac{-i}{\hbar}\right)\int_0^t\hat{H}_{ee}(t_1)dt_1
\\
\\
\,\,\,\,\,\,\,\,\,\,\,\,\,\,\,\,\,\,\,
+\left(\frac{-i}{\hbar}\right)^2
\int_0^t\hat{H}_{ee}(t_1)\int_0^{t_1}\hat{H}_{ee}(t_2)dt_2dt_1
+\cdots.
\end{array}
\end{equation}
We assume $\delta=\Delta$ for simplicity, then the above
time-evolution operator can be approximated as
\begin{equation}
\hat{U}(t)\approx\exp\left(-\frac{it}{\hbar}\hat{H}_{\rm{eff}}\right),
\end{equation}
with the effective Hamiltonian
\begin{equation}
\begin{array}{l}
\hat{H}_{\rm{eff}}=\frac{\hbar
\Omega^2}{\delta}\left[\hat{a}^\dagger\hat{a}
(\hat{\sigma}_+\hat{\sigma}_--\hat{\sigma}_-\hat{\sigma}_+)
+\hat{\sigma}_+\hat{\sigma}_-\right]
\\
\\
\,\,\,\,\,\,\,\,\,\,\,\,\,\,\,\,\,\,\,\,\,\,\,\,\,\,+ \frac{\hbar
\tilde{\Omega}^2}{\delta}
\left(\hat{b}^\dagger\hat{b}-\hat{a}^\dagger\hat{a}\right)
+\frac{\hbar\Omega\tilde{\Omega}}{\delta}\left(\hat{\sigma}_+
\hat{b} +\hat{\sigma}_-\hat{b}^\dagger\right).
\end{array}
\end{equation}
The second term in the right hand of Eq.~(10) and the terms relating
to the high orders of $\Omega/\delta$ and $\tilde{\Omega}/\delta$
were neglected, since $\Omega$, $\tilde{\Omega}\ll\delta$.
Furthermore, at the experimental temperature (e.g., $20$ mK) the
electrons are frozen well into their vibrational ground states
(about $40$~mK for the vibrational frequency $\sim10$~GHz). This
means that the excitation of the vibration of electron $e_1$ is
virtual, and thus the terms in Eq.~(12) related to
$\hat{a}^\dagger\hat{a}$ can be adiabatically eliminated. As a
consequence, the Hamiltonian in Eq.~(12) reduces to
\begin{equation}
\hat{H}_{\rm{eff}}=\frac{\hbar\Omega^2}{\delta}\hat{\sigma}_+\hat{\sigma}_-
+\frac{\hbar \tilde{\Omega}^2}{\delta}\hat{b}^\dagger\hat{b}
+\frac{\hbar
\Omega\tilde{\Omega}}{\delta}\left(\hat{\sigma}_+\hat{b}
+\hat{\sigma}_-\hat{b}^\dagger\right)
\end{equation}
and further reads (for $\Omega=\tilde{\Omega}$)
\begin{equation}
\hat{H}_{\rm{JC}}=\frac{\hbar\Omega^2}{\delta}
\left(\hat{\sigma}_+\hat{b} +\hat{\sigma}_-\hat{b}^\dagger\right)
\end{equation}
in the interaction picture. Obviously, this Hamiltonian describes a
JC-type coupling between the spin of electron $e_1$ and the orbital
motion of electron $e_2$.

Typically, the effective coupling strength can reach
$\Omega'=\Omega^2/\delta\approx2.5$~MHz for $d=10\,\mu$m,
$\nu_{1x}=10$~GHz, and $\delta=250$~MHz. With these parameters and
the Hamiltonian in E.q~(8), Fig.~3 shows numerically the occupancy
evolutions of the states $|\uparrow_1,0_1,0_2\rangle$ and
$|\downarrow_1,0_1,1_2\rangle$. Here, $|\downarrow_{j}\rangle$ and
$|\uparrow_{j}\rangle$ are the two spin states of electron $e_j$,
and $|0_j\rangle$ and $|1_j\rangle$ are the two lower vibrational
states of the electron. Obviously, the results are well agreement
with the solutions (i.e., the time-dependent occupancies of
$|\uparrow_1,0_2\rangle$ and $|\downarrow_1,1_2\rangle$) from the
Hamiltonian $\hat{H}_{\rm{JC}}$. This verifies the validity of
$\hat{H}_{\rm{JC}}$.
\begin{figure}[tbp]
\includegraphics[width=8cm]{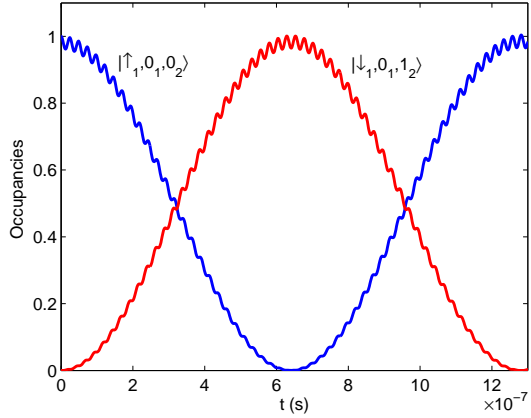}
\caption{(Color online) Numerical solutions for the Hamiltonian in
Eq.~(8): the occupancy evolutions of states
$|\uparrow_1,0_1,0_2\rangle$ (blue curve) and
$|\downarrow_1,0_1,1_2\rangle$ (red curve), with
$\tilde{\Omega}=\Omega=25$~MHz and $\delta=\Delta=250$~MHz.}
\end{figure}
The spin-orbit JC coupling (14) could be used to implement QIP
between the separately trapped electrons.
For example, by applying a current pulse with the duration
$t=\pi/(2\Omega')$ to an electrode, e.g., $\text{I}_{1}$, a
two-qubit operation $\hat{V}_{1,2}(\pi/2)
=|\downarrow_{1},0_{2}\rangle\langle\downarrow_{1},0_{2}|
-i|\downarrow_{1},1_{2}\rangle\langle\uparrow_{1},0_{2}|$ between
the electrons could be implemented. Consequently, a CNOT gate
between the qubits encoded by the electronic spins could be
implemented by the operational sequence
$\hat{C}=\hat{V}_{1,2}(\pi/2)\hat{S}_2\hat{V}_{1,2}(\pi/2)$, with
$\hat{S}_2$ being the single-electron CNOT gate operated on the
electron $e_2$. After this two-spin CNOT operation, the vibrational
motions of the trapped electrons return to their initial ground
states.

Furthermore, the mechanism used above for the distant spin-orbit
coupling can be utilized to implement an orbit-mediated spin-spin
interaction, wherein the degrees freedom of the orbits of the two
electrons are adiabatically eliminated. Indeed, by applying the
current pulses to the electrodes simultaneously, the Hamiltonian of
the individually-driven electrons reads:
\begin{equation}
\hat{H}'_{ee}=\hbar\Omega\left(e^{i\delta t}\hat{\sigma}_+\hat{a}
+e^{-i\delta t}\hat{\sigma}_-\hat{a}^\dagger\right)+\hbar
\tilde{\Omega}\left(e^{i\delta t}\hat{a}\hat{b}^\dagger+e^{-i\delta
t}\hat{a}^\dagger\hat{b}\right)+\hbar G\left(e^{i\eta
t}\hat{\tau}_+\hat{b}+e^{-i\eta t}
\hat{\tau}_-\hat{b}^\dagger\right).
\end{equation}
Here, the first and third terms describe respectively the spin-orbit
couplings of the electrons $e_1$ and $e_2$, and the second term
describes the Coulomb interaction between the electrons. $G$ and
$\eta$ are the coupling strength and the detuning between the spin
and orbital motions of electron $e_2$, respectively.
$\hat{\tau}_-=|\downarrow_2\rangle\langle\uparrow_2|$ and
$\hat{\tau}_+=|\uparrow_2\rangle\langle\downarrow_2|$ are the
corresponding spin operators of electron $e_2$. The spin-orbit
couplings, i.e., the first and third terms in the Hamiltonian, can
be realized by applying the ac currents $I_1(t)=I_1\cos(\omega_1t)$
and $I_2(t)=I_2\cos(\omega_2 t)$ to the electrodes $\text{I}_1$ and
$\text{I}_2$ respectively, with the frequencies
$\omega_1=\nu_{1x}-\nu_s+\delta$ and $\omega_2=\nu_{2x}-\nu_s+\eta$.
Here the ac currents are applied to relatively-easily satisfy the
above requirements for the detunings.

\begin{figure}[tbp]
\includegraphics[width=8cm]{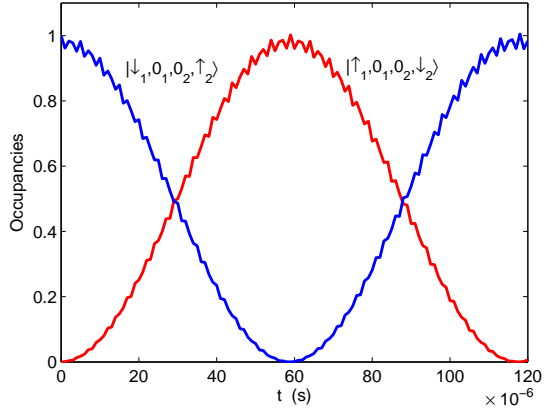}
\caption{(Color online) Numerical solutions for the Hamiltonian in
Eq.~(15): the occupancy evolutions of the states
$|\downarrow_1,0_1,0_2,\uparrow_2\rangle$ (blue curve) and
$|\uparrow_1,0_1,0_2,\downarrow_2\rangle$ (red curve), with
$\tilde{\Omega}=25$~MHz, $\Omega=2.6$~MHz, $\delta=250$~MHz, and
$\eta=\Omega^2/\delta$.}
\end{figure}
With the help of Eq.~(13), Eq.~(15) can be effectively simplified as
\begin{equation}
\hat{H}'_{ee}=\hat{H}_{\rm{eff}}+\hbar G\left(e^{i\eta
t}\hat{\tau}_+\hat{b}+e^{-i\eta t}
\hat{\tau}_-\hat{b}^\dagger\right),
\end{equation}
i.e.,
\begin{equation}
\hat{H}'_{ee}=\frac{\hbar\Omega\tilde{\Omega}}{\delta}\left(e^{i
\gamma t}\hat{\sigma}_+\hat{b}+e^{-i
t\gamma}\hat{\sigma}_-\hat{b}^\dagger\right)+\hbar
G\left(e^{i(\eta-\tilde{\Omega}^2/\delta)
t}\hat{\tau}_+\hat{b}+e^{-i(\eta-\tilde{\Omega}^2/\delta) t}
\hat{\tau}_-\hat{b}^\dagger\right)
\end{equation}
in the interaction picture, with
$\gamma=(\Omega^2-\tilde{\Omega}^2)/\delta$. We select
$G=\Omega\tilde{\Omega}/\delta$ and $\eta=\Omega^2/\delta$ for
simplicity, such that
\begin{equation}
\hat{H}'_{ee}=\hbar G\left(e^{i \gamma t}\hat{\sigma}_+\hat{b}+e^{-i
t\gamma}\hat{\sigma}_-\hat{b}^\dagger\right)+\hbar G\left(e^{i
\gamma t}\hat{\tau}_+\hat{b}+e^{-i\gamma
t}\hat{\tau}_-\hat{b}^\dagger\right).
\end{equation}
By repeating the same method for deriving the effective Hamiltonian
$\hat{H}_{\rm{eff}}$, i.e., neglecting the terms relating to the
high orders of $G/\gamma$ in the time-evolution operator and
eliminating adiabatically the terms relating to
$\hat{b}^\dagger\hat{b}$, we have
\begin{equation}
\hat{H}_{\rm{eff}}'=\frac{\hbar G^2}{\gamma}
\left(\hat{\sigma}_+\hat{\tau}_-+\hat{\sigma}_-\hat{\tau}_+\right).
\end{equation}
This is an effectively interaction between the two spins, mediated
by their no-excited orbital motions~\cite{MS}.

Numerically, for $\tilde{\Omega}\approx25$~MHz,
$\Omega\approx2.6$~MHz, and $\delta\approx250$~MHz, we have
$|\gamma|\approx2.5$~MHz, $G\approx0.26$~MHz, and
$\Omega''=|G^2/\gamma|\approx 27$~kHz. With these parameters, Fig.~4
shows numerically the time-dependent occupancies of
$|\downarrow_1,0_1,0_2,\uparrow_2\rangle$ and
$|\uparrow_1,0_1,0_2,\downarrow_2\rangle$ from the Hamiltonian in
Eq. (15). This provides the validity of the simplified Hamiltonian
in Eq. (19). Obviously, the present orbit-mediated spin-spin
coupling is significantly weaker than the above spin-orbit JC
coupling (14) between the electrons, but still stronger than the
directly magnetic dipole-dipole coupling (which is estimated as
$\sim10^{-3}$~Hz for the same distance) between the spins. Since the
coherence time of the spin qubit is very long (e.g., could be up to
minutes~\cite{Spin}), the orbit-mediated spin-spin coupling
demonstrated above could be utilized to generate the spins
entanglement and thus implement the desirable QIP.

Finally, we would like to emphasize that, the considered double-trap
configuration shown in Fig.~2 seems similarly to that of the recent
ion-trap experiments~\cite{Ion-1,Ion-2}. There, two ions are
confined in two potential wells separated by $40\,\mu$m~\cite{Ion-1}
(or $54\,\mu$m~\cite{Ion-2}), and the ion-ion vibrational coupling $
\hat{H}_{ii}=\hbar\tilde{\Omega}[\exp(i\Delta
t)\hat{a}\hat{b}^\dagger+\exp(-i\Delta t)\hat{a}^\dagger\hat{b}]$ is
achieved up to $\tilde{\Omega}\approx10$~kHz~\cite{Ion-1} (or
$\tilde{\Omega}\approx7$~kHz~\cite{Ion-2}). The coupling between the
ions was manipulated tunably by controlling the potential wells (via
sweeping the voltages on the relevant electrodes) to adiabatically
tune the oscillators into or out of resonance, i.e., $\Delta=0$ or
$\Delta\gg\tilde{\Omega}$~\cite{Miao-i}, respectively.
Instead, in the present proposal we suggested a JC-type coupling
(and consequently an orbit-mediated spin-spin coupling) between the
two separated electrons. Therefore, the operational steps for
implementing the QIP should be relatively simple. More
interestingly, here the electron-electron coupling strength
$\tilde{\Omega}$ is significantly stronger (about $10^3$ times) than
that between the trapped ions~(e.g, $^{9}\rm {Be}^+$~\cite{Ion-1}),
since the mass of electron is much smaller than that of the ions.

\section{conclusion}
We have suggested an approach to implement the QIP with electronic
spins on liquid helium. Two long-lived spin states of the trapped
electron were encoded as a qubit, and the strong Coulomb interaction
between the electrons was utilized as the data bus. The spin-orbit
JC coupling between the spin of an electron and the vibrational
motion of another distant electron is generated by designing a
virtual excitation of the electronic vibration. Such a distant
spin-orbit interaction is further utilized to realize an
orbit-mediated spin-spin coupling and implement the desirable
quantum gates.

Compared with the ions in the Paul traps, here a feature is that the
mass of the electron is much smaller than that of ions, and thus a
strong Coulomb coupling up to 25 MHz between the electrons could
reached for a distance of $d=10\,\mu$m.
Finally, the construction suggested here for implementing quantum
computation with trapped electrons on the liquid helium should be
scalable, and hopefully be feasible with current micro-scale
technique.

{\bf Acknowledgements}: This work was partly supported by the
National Natural Science Foundation of China Grants No. 11204249,
11147116, 11174373, and 90921010, the Major State Basic Research
Development Program of China Grant No. 2010CB923104, and the open
project of State Key Laboratory of Functional Materials for
Informatics.
\\


\begin{thebibliography}{99}
\bibitem{ESR}
Alternatively, the single-bit operations could be implemented by
applying an ac current $I(t)$ (with frequency $\omega=\nu_s$ and
phase $\theta$ on the electrode $\rm{I}_{2}$, see, Fig.~2), which
generates a magnetic field $B_{z}\approx\mu_0I(t)d/[2\pi (d^2+h^2)]$
along the $z$-direction to excite resonantly the spin of electron
$e_1$.
\end{thebibliography}
\end{document}